\documentclass[useAMS,usenatbib]{mn2e}
\usepackage{graphicx,subfigure,amsmath,wasysym, amssymb}

\title{Dust and Ionized Gas Association in E/S0 Galaxies with Dust Lanes: Clues to their Origin}

\author[Ido Finkelman et al.]{Ido Finkelman$^{1}$\thanks{E-mail:ido@wise.tau.ac.il (IF)}, Noah Brosch$^{1}$, Jos\'{e} G. Funes S.J.$^{2}$, Sudhanshu Barway$^{3}$,
\newauthor Alexei Kniazev$^{3,4}$, Petri V\"{a}is\"{a}nen$^{3,4}$\\
$^{1}$The Wise Observatory and the School of Physics and
Astronomy, the Raymond and  Beverly Sackler Faculty of Exact
Sciences, \\ Tel Aviv University, Tel Aviv 69978, Israel\\
$^{2}$Vatican Observatory, V-00120 Vatican City State, Italy\\ 
$^{3}$South African Astronomical Observatory, PO Box 9, 7935 Observatory, Cape Town, South Africa \\
$^{4}$Southern African Large Telescope Foundation, PO Box 9, 7935 Observatory, Cape Town, South Africa \\}

\date{Accepted 2012 February 7.  Received 2012 February 6; in original form 2011 September 18}
      
\pagerange{\pageref{firstpage}--\pageref{lastpage}} 

\begin{document}

\maketitle

\label{firstpage}

\begin{abstract}
We present results from an on-going programme to study the dust and ionized gas in E/S0 galaxies with dust lanes.
Our data, together with results from previous studies of E/S0 galaxies, are used to demonstrate the tight relationship between these two components. This relationship is discussed in light of our current understanding of the nature and origin of the interstellar medium (ISM), and in particular in the context of the interplay between the different multi-temperature components.
We show that focusing on dust obscured regions as tracers of the ISM, and on their properties, serves as independent evidence for the external origin of the dust and ionized gas.
\end{abstract}


\section{Introduction}
How do E/S0 galaxies acquire their ISM? What are the processes that regulate the evolution of the ISM? 
Posing such questions, and considering their answers, modifies our traditional view of elliptical galaxies as simple and dull objects while challenging our understanding of the classical Hubble classification scheme.

The growing amount of observational data accumulated by recent multi-wavelength surveys implies that there is no simple answer to the longstanding origin question (de Zeeuw et al.\ 2002; Rampazzo et al.\ 2005; Cappellari et al.\ 2011). The general picture that emerges from these studies is that both accretion and merger events and internal production could have contributed to some extent to the overall ISM content in E/S0 galaxies. The relative contribution of external versus internal sources does not seem to follow a general rule but probably varies between objects.

To delve into the properties of the ISM in E/S0 galaxies much attention has been paid to spectroscopic analysis of the dynamics and chemical abundance of the gas and stars (Sarzi et al.\ 2006; Annibali et al.\ 2010; Young et al.\ 2011).
Although highly informative, the limited spectral coverage of these investigations, together with our current limited knowledge of the interplay between different gas phases, tend to lead to an indecisive conclusion regarding the true nature of the ISM. 
Moreover, although it was found that the motion and orientation of the gas in many systems are decoupled from the stellar rotation, the overall distribution of the kinematical misalignment between stars and gas in E/S0 galaxies is inconsistent with a purely external origin (Sarzi et al.\ 2006).

The detection of unusual optical signatures in such systems could provide more definitive and direct indications to the dominant source of the ISM. This is particularly true for a subclass of galaxies where dust is easily recognized by its light obscuration which produces well-defined dark lanes. 
A significant fraction of these galaxies exhibit morphological disturbances, such as shells and tidal features, indicative of recent mergers or accretion events (van Dokkum 2005; Tal et al.\ 2009; Kaviraj et al.\ 2011).
In fact, when a single stellar population evolves the ejected material is expected to spread evenly throughout the galaxy,  
whereas the spatial distribution and orientation of the optical dust features are frequently misaligned with those of the stars (van Dokkum \& Franx  1995; Tran et al.\ 2001; Krips et al.\ 2003; Martel et al. 2004; Quillen 2006; Finkelman et al.\ 2010a).
Furthermore, the dust content of these galaxies, estimated by measuring total optical extinction, is usually found to be several times larger than expected to be supplied by internal processes, such as stellar mass loss (Patil et al.\ 2007). 

Studying the interplay between the various multi-temperature components of the ISM in E/S0 galaxies with dust lanes provides also vital clues to the external origin of the dust.
Large-scale dust structures are virtually always accompanied by excited or ionized gas, whose angular momenta are typically relatively high and often orthogonal to those of the stars (Bertola 1987; Kormendy \& Djorgovski 1989; Caon, Pastoriza \& Macchetto 2001; Krajnovic et al.\ 2008; Davis et al.\ 2011). 
E/S0 galaxies with dust lanes show also high detection rates of cold molecular gas (Sage \& Galletta 1993; Wang, Kenney \& Ishizuki 1992; Combes, Young \& Bureau 2007; Krips et al.\ 2010; Young et al.\ 2011), whereas large-scale dust features are also often associated with HI gas in discs rotating at random orientations with respect to the optical major axis (Kormendy \& Djorgovski 1989; Oosterloo et al.\ 2002; Leeuw et al.\ 2007). Perhaps more intriguing is the detection of significant dust reservoirs in X-ray bright E/S0 galaxies, where sputtering by hot particles is expected to erode and destroy the dust grains on short time scales of $\sim$100 Myr.

To tie these pieces of information Sparks et al.\ (1989) and de Jong et al.\ (1990) independently proposed the `evaporation flow' scenario, in which cold dust and gas are externally acquired by an elliptical galaxy during a gravitational interaction with a gas-rich galaxy, and then heated due to the thermal interaction with the ambient hot plasma.
As a result, the interacting hot gas locally cools to its warm (ionized) phase, whereas the dust is partially shielded within the gradually evaporating dense molecular clouds. 
This scenario accounts also for the tight spatial correlation between dust and ionized gas that is often observed along filaments, or cooling flows, associated with the presence of hot X-ray emitting coronal gas (Macchetto et al.\ 1996; Trinchieri \& Goudfrooij 2002; Fabian et al.\ 2003; sparks et al.\ 2004). 
Unlike cold dust and molecular gas, atomic hydrogen is not expected to survive in the presence of hot X-ray gas, which explains the HI deficiency in the central parts of E/S0 galaxies with dust lanes (Oosterloo et al. 2002; Leeuw et al.\ 2008).

This paper summarizes the results of an ongoing study of a sample of E/S0 galaxies where dust lanes have been reported in the literature. Our results, combined with those from similar studies, are used to demonstrate that the properties of the observed dust strongly support the `evaporation flow' scenario sketched above.

The paper is organized as follows: Section \ref{S:Obs_and_Red}  
gives a short description of the sample, observations and data reduction; Section \ref{S:analysis} discusses the data analysis and results; the results are discussed in Section \ref{S:discuss} and our conclusions are summarized in Section \ref{S:conclusions}.
We shall assume throughout the paper standard cosmology with $H_0=73$ km s$^{-1}$ Mpc$^{-1}$, $\Omega_m=0.27$ and $\Omega_\Lambda=0.73$.

\section{Observations and data reduction}
\label{S:Obs_and_Red}
This is the second paper in our survey of the ISM in a distinct morphological subclass of early-type galaxies (see Finkelman et al.\ 2010a).
Our sample consists primarily of galaxies identified by early photographic catalogues as elliptical stellar bodies crossed by extended dust lanes (Hawarden et al.\ 1981, Ebneter \& Balick 1985; Bertola 1987; V\'{e}ron-Cetty \& V\'{e}ron 1988).
Since the dust lanes presumably represent highly-inclined structures, the optically-obscured elliptical galaxies were often confused with lenticular galaxies, although (almost) no trace of a stellar disc was later identified in their CCD images. 
However, it seems at present, in many aspects, less important whether to call such objects elliptical or lenticular galaxies; we therefore refer to them as transitional dusty `E/S0' galaxies. 

Other objects are selected from more recent studies of E/S0 galaxies with dust lanes, including the search for cold gas (Gregorini, Messina \& Vettolani 1989; M\"{o}llenhoff, Hummel \& Bender 1992; Wang, Kenney \& Ishizuki 1992) and the determination of the extragalactic dust extinction law in the dark lanes (Goudfrooij et al.\ 1994b; Patil et al.\ 2007; Finkelman et al.\ 2008; Finkelman et al.\ 2010b).
The sample galaxies are listed in Table \ref{t:Obs} with their coordinates, morphological classification, integrated blue luminosity, heliocentric velocity and optical size taken from NED.

As part of our on-going programme we observed 20 more E/S0 galaxies with broad-band and H$\alpha$ narrow-band filters, adding to the sample discussed in Finkelman et al.\ (2010a). The observations were conducted during 2009 and 2010 with the 1.8-m Vatican Advanced Technology Telescope at the Mt.\ Graham International Observatory (MGIO), the 1.9-m telescope at the South African Astronomical Observatory (SAAO) and the Wise Observatory 1-m telescope.
Optical and near-IR images of NGC 5128 were taken with the CTIO 0.9-m telescope in May 2001 and with the infrared survey facility 1.4-m telescope in SAAO in April 2010, respectively.
Typical exposure times used for the galaxies in our sample were 10 min for the optical and near-IR filters and 20 min for the H$\alpha$ narrow-band filters.

Detailed descriptions of data reduction steps, flux calibration and error propagation are available in Finkelman et al.\ (2010a).
Below we briefly summarize these steps.

\begin{table*}
 \centering
  \caption{Global parameters for galaxies in our sample.
  \label{t:Obs}}
\begin{tabular}{|lcrlrrcl|}
\hline
Object      & RA       & \multicolumn{1}{c}{DEC}       & Morph.\       & \multicolumn{1}{c}{B$^0_T$} & \multicolumn{1}{c}{v$_{Helio}$}  & Size & Observatory \\
{}          & (J2000.0)& (J2000.0) &(NED)&   (mag)    &   (km/s)     & (arcmin)\\
\hline 
NGC 524      & 01:24:48 & +09:32:20 & SA0           &  11.3   &    2379      & 2.8x2.8 & MGIO\\
NGC 984      & 02:34:43 & +23:23:47 & SA0           &  13.8   &    4352      & 3.0x2.0 & MGIO\\
NGC 1172     & 03:01:36 & -14:50:12 & E             &  13.3   &    1669      & 2.3x1.8 & MGIO\\
NGC 1439     & 03:44:50 & -21:55:12 & E             &  13.0   &    1670      & 2.5x2.3 & WO \\
NGC 1947     & 05:26:48 & -63:45:36 & S0 pec        &  11.7   &    1100      & 3.0x2.6 & WO \\
ESO 159-G019 & 05:33:11 & -52:38:31 & S0/a          &  14.4   &    4338      & 1.5x0.8 & SAAO\\
NGC 2076     & 05:46:47 & -16:46:57 & S0            &  14.0   &    2142      & 2.2x1.3 & SAAO\\
ESO 087-G028 & 06:33:19 & -62:59:39 & S0            &  14.5   &    8444      & 1.1x0.7 & SAAO\\
NGC 2907     & 09:31:37 & -16:44:05 & SAa LINER?    &  12.7   &    2090      & 1.8x1.1 & SAAO\\ 
NGC 2911     & 09:33:46 & +10:09:09 & SA0 Sy LINER  &  12.5   &    3183      & 4.1x3.2 & WO \\
NGC 3302     & 10:35:47 & -32:21:31 & SA0           &  13.5   &    4075      & 1.7x1.2 & SAAO\\
NGC 3489     & 11:00:19 & +13:54:04 & SAB0 Sy2      &  11.1   &    677       & 3.5x2.0 & WO \\
NGC 3497     & 11:07:18 & -19:28:18 & SA0           &  13.0   &    3672      & 2.6x1.4 & SAAO\\
NGC 4753     & 12:52:22 & -01:11:59 & I0            &  10.9   &    1239      & 6.0x2.8 & WO \\
NGC 5128     & 13:25:28 & -43:01:09 & S0 pec Sy2    &  7.8    &    547       & 25.7x20.0 & CTIO, SAAO\\
NGC 5266     & 13:43:02 & -48:10:10 & SA0 LINER     &  12.1   &    3002      & 3.2x2.1 & SAAO\\
IC 4320      & 13:44:04 & -27:13:54 & S0            &  14.2   &    6805      & 1.0x1.0 & SAAO\\
ESO 384-G012 & 13:55:34 & -33:54:01 & S0 pec        &  14.3   &    4577      & 1.1x0.9 & SAAO\\
NGC 5525     & 14:15:39 & +14:16:57 & S0            &  13.8   &    5553      & 1.4x0.9 & WO \\
NGC 7722     & 23:38:41 & +15:57:17 & S0/a          &  13.5   &    4026      & 1.7x1.2 & WO \\
\hline
\end{tabular}
\end{table*}

Image reduction was performed with standard tasks within {\small IRAF}\footnote{{\small IRAF} is distributed by the National Optical Astronomy Observatories (NOAO), which is operated by the Association of Universities, Inc. (AURA) under co-operative agreement with the National Science Foundation}. These include bias subtraction, overscan subtraction and flatfield correction. 
Cosmic rays events were removed from single CCD exposures by using the L.A.Cosmic task in IRAF (van Dokkum 2001) while CCD hot pixels were removed with the FIXPIX task in {\small IRAF} using an appropriate mask. 

The reduced images are background-subtracted and geometrically aligned by measuring centroids of several common stars in the galaxy frames. This alignment procedure involves IRAF tasks for scaling, translation and rotation of the images, so that a small amount of blurring is introduced affecting the accuracy to be better than a few tenths of an arcsec. 
We observed each galaxy using two filters, a narrow-band filter which covers the rest-frame H$\alpha$+[NII] emission, and a broad R-band filter.
The H$\alpha$ images include photons from the H$\alpha$ and the [NII] lines and from the continuum. Since we are interested in the H$\alpha$ line,
the R-band image is scaled to match the intensity of the stellar continuum in the narrow-band image and is subtracted from the narrow-band image.
Finally, the measured H$\alpha$+[NII] counts are converted into physical units by observing standard stars (Landolt 1992). 

Our analysis also include archival images taken from the Two Micron All Sky Survey (2MASS), the Wide-field Infrared Survey Explorer (WISE) and the Infrared Astronomical Satellite (IRAS). All images were reduced and flux-calibrated using the standard pipelines.
We note that the WISE Preliminary Release covers only about half of the sky and therefore does not include many of our objects.

\section{Analysis and results}
\label{S:analysis}
\subsection{Dust grain properties}
Dark lanes are produced by the absorption and scattering of optical light by dust grains in the dusty structure. The amount of extinction and its effect on observed colours depends strongly on the size distribution, structure and chemical composition of the grains. 
Fitting the unextinguished parts of an E/S0 galaxy with an underlying smooth light distribution 
allows the estimation the dust extinction in these regions. This is done by fitting the galaxies with elliptical isophotes using the ISOPHOTE package in {\small IRAF} and subtracting the observed galaxy image from the dust-free model to measure the extinction. 
Applying this method is useful for characterizing the dust properties and estimating the dust mass.

The wavelength dependence of extinction in almost all E/S0 galaxies with dust lanes follows closely the standard Galactic extinction law, showing only small departures (Goudfrooij et al.\ 1994b; Patil et al.\ 2007; Finkelman et al.\ 2008, 2010a, 2010b).
For those galaxies in our sample not included in previous similar studies we validated this result by measuring the extinction of light in the optical bands, with the exception of NGC 5128 which was observed in the near-IR (see also Kainulainen et al.\ 2009). Considering the well-studied properties of Galactic dust grains and their similarity with dust grains in dust lanes we conclude that the typical size of the (large form of) dust grains in our sample galaxies is about $\sim$0.1 $\mu$m (see Draine \& Lee 1984; Casuso \& Beckman 2010).
The energy absorbed by these dust grains is re-emitted in far-IR and sub-mm wavelengths.

\begin{table*}
 \centering
  \caption{Dust mass and ionized gas mass derived from optical images. Masses are given in Solar units.
  \label{t:Mass}}
\begin{tabular}{|llll|}
\hline
Object        & $ \mbox{Log}\left( \frac{M_{\mbox{dust,opt}}}{M_\odot}\right) $  &  $ \mbox{Log}\left( \frac{M_{\mbox{dust,IRAS}}}{M_\odot}\right) $ & $\mbox{Log} \left(\frac{M_{\mbox{HII}}}{M_ \odot}\right)$  \\
\hline 
NGC 524$^{ab}$ & $4.58\pm0.14$ & $5.58\pm0.21$ & $3.60$\\
NGC 984       & $5.46\pm0.01$ & $4.25\pm0.47$ & $5.28\pm0.25$ \\
NGC 1172$^{ac}$ & $3.83\pm0.18$ & $<5.13$* & $3.42$\\
NGC 1439$^a$ & $5.10\pm0.19$ & $<5.77$* & $<3.48$\\
NGC 1947$^d$ & $5.01\pm0.01$ & $5.46\pm0.06$ & $3.47\pm0.14$\\
ESO 159-G019 & $5.62\pm0.01$ & $6.52\pm0.07$ & $5.41\pm0.05$ \\
NGC 2076$^e$ & $6.18$ & $6.59\pm0.08$ & $4.53\pm0.09$\\
ESO 087-G028 & $6.26\pm0.01$ & $<7.02$* & $<4.69$\\
NGC 2907     & $6.13\pm0.01$ & $5.31\pm0.22$ & $<3.83$\\ 
NGC 2911     & $5.80\pm0.02$ & $5.76\pm0.26$ & $4.67\pm0.22$\\
NGC 3302$^d$ & $5.55\pm0.02$ & $6.00\pm0.23$ & $<4.65$  \\
NGC 3489     & $4.00\pm0.01$ & - & $3.71\pm0.18$ \\
NGC 3497$^f$ & $6.03\pm0.03$  & $6.34\pm0.29$ & $ 5.13\pm0.11$  \\
NGC 4753$^{dg}$& $5.20\pm0.01$ & $5.96\pm0.09$ & $5.43$\\
NGC 5128     & $6.58\pm0.08$ & $6.03\pm0.07$ & $4.99\pm0.01$\\
NGC 5266$^d$ & $5.04\pm0.03$ & $6.16\pm0.02$ & $5.29\pm0.13$\\
IC 4320      & $6.15\pm0.01$ & $<6.80$* & $<4.85$  \\
ESO 384-G012 & $5.64\pm0.02$ & $4.48\pm0.88$ & $<4.37$ \\
NGC 5525     & $6.23\pm0.01$ & - & $<4.75$\\
NGC 7722     & $6.42\pm0.01$ & $6.64\pm0.15$ & $5.07\pm0.19$\\
\hline
\end{tabular}
\begin{minipage}[]{17.5cm}
\begin{footnotesize}
*Upper limits assuming $\mbox{T}=20\mbox{K}$\\
{\bf References:} a - Patil et al.\ (2007); b - Sarzi et al.\ (2006); c - Macchetto e tal.\ (1996); d - Finkelman et al.\ (2010b); e - Sahu et al.\ (1999); f - Finkelman et al.\ (2008); g - Dewangan et al.\ (1999).
\end{footnotesize}
\end{minipage}
\end{table*}
\subsection{Dust mass}
\subsubsection{Optical extinction}
The dust mass is calculated by integrating the dust column density $\Sigma_d$ over the image areas S occupied by dust lanes.
The dust column density can be estimated from the total extinction values by assuming a chemical composition of the extragalactic dust grains similar to that of the dust in the Galaxy.
While the details of these calculations are described in Finkelman et al.\ (2008), a useful approximation of the dust mass can be given as
\begin{equation}
M_d=\mbox{S}{\times}\Sigma_d=\mbox{S}{\times}l_d{\times}n_H{\times}\int\limits_{a_-}^{a_+}\frac{4}{3}{\pi}a^3{\rho}_df\left(a\right)da .
\end{equation}
where $a$ is the grain radius; $f\left( a\right)$ represents the size distribution of dust grains; $a_-$ and $a_+$ represent the lower and upper cutoffs of the size distribution, respectively; ${\rho}_d$ is the specific grain densities taken to be $\sim 3$ gr cm$^{-3}$ (Draine \& Lee 1984); $l_d$ represents the dust column length along the line of sight; and $n_H$ is the hydrogen number density.
Note that we make no assumption about the gas-to-dust relationship to estimate the hydrogen column density. Instead, the value of the product $l_d \times n_H$ is inferred by measuring the total extinction and calculating the extinction efficiency (Finkelman et al.\ 2008). 
This measurement provides only a lower limit to the true dust content of the host galaxies since our calculations assume that the dust forms a foreground screen for the galaxy. 

\subsubsection{Thermal emission}
The average temperature of the dust in each galaxy can be estimated by fitting modified blackbody functions to the IR data.
In the wavelength range $\lambda \gtrsim 20\mu$m, the emission of isothermal dust grains can be modeled with a dust emissivity law given by $F_\nu \propto B_\nu (T_d) \nu ^{\beta}$, where $F_{\nu}$ and $B_{\nu}(T_d)$ are the flux density and the Planck function for the temperature $T_d$ at frequency $\nu$, respectively. The dust emissivity index $\beta$ is generally taken to be between 1.0 and 2.0 (Draine \& Lee 1984; Genzel \& Cesarsky 2000). 

Given a $\beta\approx 1.0$ emissivity law, the dust grain temperature can be calculated from the IRAS flux densities at 60 $\mu$m and 100 $\mu$m using $T_d=\left( \frac{S_{60}}{S_{100}}\right)^{0.4}$  (Young et al.\ 1989).
While cold dust was already known to reside in normal, inactive spiral galaxies, IRAS observations established the presence of `warm' ($\sim$30-40 K) dust which dominates the 60 $\mu$m and 100 $\mu$m bands.
However, IRAS did not cover the spectral range beyond 100 $\mu$m and was therefore insensitive to dust colder than $\sim$20 K.

The far-IR emission of E/S0 galaxies can be used to estimate their dust content independently of the optical extinction and reddening measurements.
Following the analysis of Hildebrand (1983), the dust mass is given by 
\begin{equation}
\label{eq:massIR}
M_{\mbox{dust,IRAS}}=\frac{4}{3} a \rho _d D^2 \frac{F_{\nu}}{Q_{\nu}B_{\nu}(T_d)}
\end{equation}
where $D$ is the distance of the galaxy in Mpc and Q$_{\nu}$ is the grain emissivity taken from Draine (1985).  
Table \ref{t:Mass} lists the estimated dust mass from the total optical extinction and the estimated dust mass based on IRAS flux densities taken from the catalog of Knapp et al.\ (1989) for bright early-type galaxies.

Recent observations of early-type galaxies have detected also a significant emission in the near- and mid-IR wavelengths. The emission in these wavelengths is believed to be produced by very small dust grains and PAH molecules (Ferrari et al.\ 2002; Xilouris et al.\ 2004). 
In particular, the emission at $\sim$25-60 $\mu$m is probably produced by the transient heating of $\sim$0.001 $\mu$m dust grains to high temperatures (Li \& Draine 2001). While the exact treatment of this emission is complex, it can be approximated as continuous thermal emission of `hot dust' of $\sim$200 K (see Ferrari et al.\ 2002).

\begin{figure}
  \centering
    \includegraphics[width=8cm]{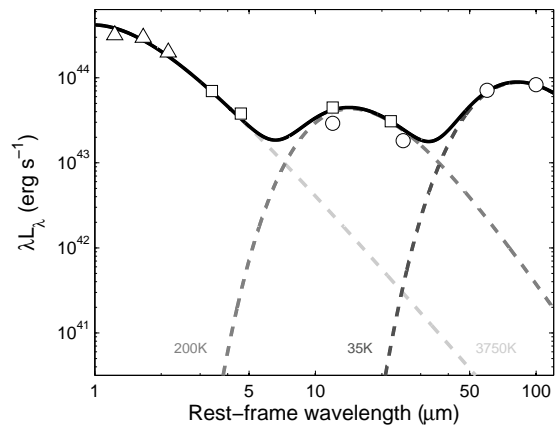}
  \caption{Spectral energy distribution (SED) of NGC 5128. Triangles, squares and circles represent 2MASS, WISE and IRAS data, respectively. The SED is well reproduced by the thermal emission of a dominant stellar population of spectral type M0, warm dust at 35 K and hot dust at 200 K.  \label{f:BBcurves} }
\end{figure}

To study the hot dust we use publicly available near-IR data taken with 2MASS and mid-IR data taken with WISE to build the spectral energy distribution (SED) and then follow the analysis by Ferrari et al.\ (2002). 
Since the hot dust thermal emission typically peaks at around 10 $\mu$m, and since up to $\sim$15 $\mu$m it includes a strong contribution of late-type stars and possibly of silicate or PAH bands (Boselli et al.\ 1998; Madden et al.\ 1999), measurements 
at longer wavelengths would serve as a more sensitive probe of the hot dust.
We therefore measure the near- and mid-IR light within an elliptical aperture enclosing the apparent emission at 22 $\mu$m (W4 - the reddest channel of WISE).
The aperture limits are set to include resolution elements with emission $\sim$1-$\sigma$ above the background level. For galaxies with no apparent extended emission structure, a circular aperture with a radius twice the FWHM seeing size around the galactic centre is examined in an attempt to detect faint dust emission.

Measuring the hot dust emission requires first to estimate the relative contribution of the Rayleigh-Jeans tail of the old stellar population.
To this purpose we match the SED with blackbody curves of three different temperatures (T=3750 K, T=4000 K and T=4600 K) that correspond to stellar populations of spectral types M0, K7 and K4. 
Note that optical images of our sample galaxies suffer significant internal extinction and are therefore not used to construct the SED.
To model the hot dust emission and determine its temperature we add a second component with a modified blackbody curve of $T_d$ in the range 40 K to 500 K. As a representative case, we show in Fig.\ \ref{f:BBcurves} the best-fit SED of NGC 5128. Fig.\ \ref{f:BBcurves} includes also a third component of a modified blackbody curve representing the warm dust component detected by IRAS. 
Although the warm dust component seems to contribute to some extent to the overall 22 $\mu$m emission of NGC 5128, we note that the IRAS spatial resolution is typically several times larger than the size of our galaxies, thus not accounted for by our modeling.

\begin{figure}
  \centering
    \includegraphics[width=8cm]{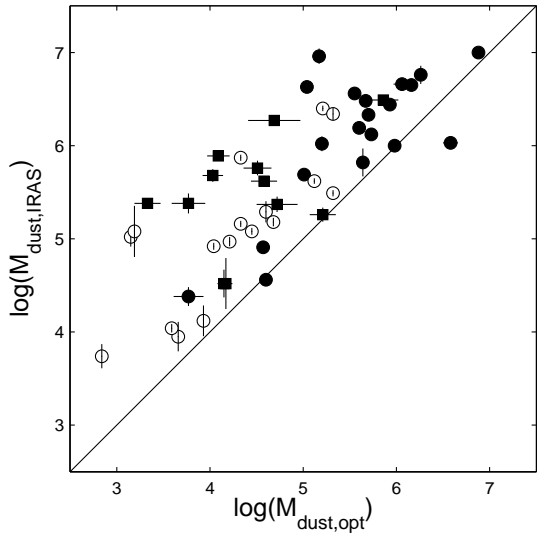}
  \caption{Dust mass derived using IRAS data versus dust mass derived using optical extinction and reddening. Our sample galaxies are represented by filled circles, open circles are from Goudfrooij et al.\ (1994a, 1994b) and filled squares are from Patil et al.\ (2007). Dust masses are given in Solar units.
  The drawn solid line is for equal masses.     \label{f:ColdDust}}
\end{figure}

We use the flux density at 22 $\mu$m, corrected by subtracting the stellar component, to estimate the hot dust mass $M_{\mbox{dust,WISE}}$ for each galaxy. This is done by using eq.\ \ref{eq:massIR} under the assumption that the quantity $a{\rho}_d/Q_{\nu}$ is independent of $a$ for $\lambda \gg a$ (see Hildebrand 1983; Draine \& Lee 1984).
The emissivity of the small grains was adjusted to be consistent with $\beta \approx 1$, as assumed above (see also Bianchi, Davis \& Alton 1999; Bendo et al.\ 2003; da Cunha, Charlot \& Elbaz 2008).

The dust masses estimated based on the mid-IR data are listed in Table \ref{t:hot} along with the total luminosity at 22 $\mu$m and the best-fit hot dust temperature.
In addition, we compare the dust masses derived using reddening and optical extinction with those using IRAS flux densities and with those derived using WISE flux densities and plot our results in Figs. \ref{f:ColdDust} and \ref{f:DustWarm}, respectively.
To improve the coverage of these diagrams, we include also data from previous studies of similar objects (Goudfrooij et al.\ 1994; Ferrari et al.\ 1999;  Martel et al.\ 2004; Patil et al.\ 2007; Finkelman et al.\ 2008, 2010a, 2010b). 

\subsection{Ionized gas mass}
The H$\alpha$+[NII] flux of each galaxy is measured following the procedure detailed in Finkelman et al.\ (2010a). 
Assuming case B recombination (Osterbrock 1989), the H$\alpha$ luminosity $L_{\mbox{H}\alpha}$ can be used to roughly estimate the mass of ionized hydrogen.
For a given electron temperature and density this mass can be written as:
\begin{equation}
M_{{\mbox{HII}}}  = \left( L_{\mbox{H}\alpha} m_H / n_e\right) / \left( 4 \pi j_{\mbox{H}\alpha} / n_e n_p \right) 
\end{equation}
where $m_H$ is the mass of the hydrogen atom; $n_e$ and $n_p$ are the number of electrons and protons per $\mbox{cm}^{3}$ and $j_{\mbox{H}\alpha}$ is the emission coefficient of the H$\alpha$ line. Assuming an electron temperature of $\sim10^4$ K the expression above can be simplified as:
\begin{equation}
M_{{\mbox{HII}}} = 2.33\times 10^3 \, \left(\dfrac{L_{\mbox{H}\alpha}}{10^{39}\mbox{ erg s}^{-1}}\right) \left( \dfrac{10^3 \mbox{ cm}^{-3}}{n_e} \right) M_\odot.
\end{equation}
We estimate the masses assuming a typical electron density of  $\sim 10^3$ cm$^{-3}$ (see Goudfrooij et al.\ 1994a) and list the results in Table \ref{t:Mass}. 

\begin{figure}
  \centering
    \includegraphics[width=8cm]{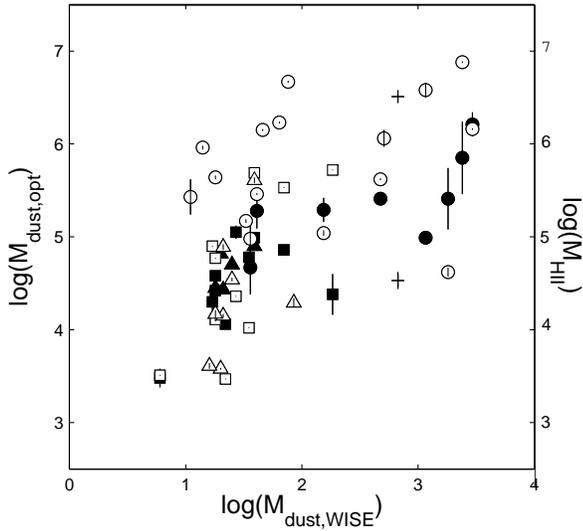}
  \caption{The relation between dust mass derived using optical extinction and dust mass derived using WISE data (open symbols) and the relation between ionized gas mass and dust mass derived using WISE data (filled symbols). Circles represent our data; squares, triangles and `+' signs represent data from Goudfrooij et al.\ (1994a, 1994b), Ferrari et al.\ (1999) and Dewangan et al.\ (1999), respectively. Masses are given in Solar units. The hot dust content tends to be more massive with increasing optical dust mass; a Spearman rank test shows that the null hypothesis of no correlation has a probability of only about 1\%.
    \label{f:DustWarm}}
\end{figure}

\begin{figure}
  \centering
    \includegraphics[width=8cm]{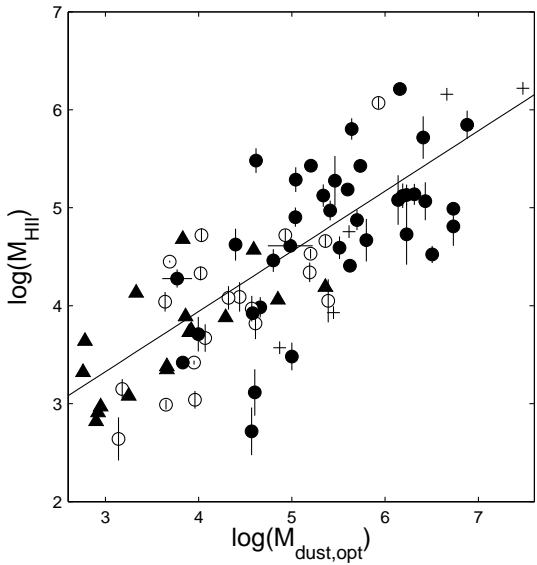}
  \caption{HII mass versus dust mass. Our sample galaxies are represented by filled circles, open circles are from Goudfrooij et al.\ (1994a, 1994b), filled triangles are from Macchetto et al.\ (1996) and `+' signs are from Martel et al.\ (2004).  Masses are given in Solar units. The dashed line is the fitted linear trend with a slope of $\sim0.8$. \label{f:DustHa}}
\end{figure}

To properly derive $L_{\mbox{H}\alpha}$ requires measuring also the ratios of the [NII] and H$\alpha$ lines, which are not available here. The ionized hydrogen masses are therefore upper limits, while the true values are likely lower by a factor of $\sim$2-3 (see Goudfrooij et al.\ 1994a; Macchetto et al.\ 1996; Finkelman et al.\ 2010a). We compare the ionized gas masses with the dust masses derived using optical extinction and with those derived using WISE flux densities and plot our results in Figs.\ \ref{f:DustWarm} and \ref{f:DustHa}, respectively. Note that the H$\alpha$ luminosity values are not corrected for extinction.

\begin{table}
 \centering
  \caption{Hot dust properties including the best-fit temperature, total luminosity in the WISE 22$\mu$m (W4) band and derived hot dust mass.  
The typical errors on luminosity and dust mass are estimated to be less than 5\%.
The number in parenthesis represent the relative contribution of dust emission to the 22$\mu$m flux.
The table includes galaxies from our sample and similar studies for which WISE data are currently available.
  \label{t:hot}}
\begin{tabular}{|lcrr|}
\hline
Object        & $T_d$       & \multicolumn{1}{c}{$L \left( 22 \mu \mbox{m} \right) $}  & \multicolumn{1}{c}{$ M_{\mbox{dust,WISE}}$}   \\
{}	      & ($^\circ$K) & \multicolumn{1}{c}{($10^{30}$ erg s$^{-1}$)}             & \multicolumn{1}{c}{($M_\odot$)} \\
\hline 
NGC 662  & 180 & 14.27 (1.00)& 2930 \\
NGC 984  & 220 & 0.40 (0.84) & 41 \\
NGC 1199 & 160 & 0.22 (0.78) & 73 \\
NGC 1407$^a$ & 200 & 0.16 (0.66) & 22 \\
NGC 1439 & 180 & 0.05 (0.77) & 11 \\
NGC 1600$^a$ & 200 & 0.28 (0.55) & 39 \\
ESO 159-G012 & 220 & 4.50 (0.99) & 480 \\
NGC 2076$^b$ & 200 & 4.75 (0.99) & 670 \\
NGC 2534 & 240 & 0.45 (0.93) & 36 \\
NGC 3136$^a$ & 200 & 0.13 (0.74) & 18 \\
IC 3370$^a$  & 200 & 0.50 (0.85) & 70 \\
NGC 4589$^a$ & 200 & 0.12 (0.68) & 17 \\
NGC 4696$^a$ & 200 & 0.28 (0.69) & 39 \\
NGC 5018$^a$ & 180 & 0.90 (0.86) & 180 \\
NGC 5044$^a$ & 200 & 0.19 (0.70) & 27 \\
NGC 5128 & 200 & 8.52 (0.99) & 1160 \\
NGC 5266 & 220 & 1.49 (0.89) & 150 \\
IC 4320  & 220 & 0.46 (0.82) & 48 \\
ESO 384-G012 & 240 & 0.23 (0.88) & 18 \\
NGC 5525 & 240 & 0.81 (0.88) & 64 \\
NGC 5576$^a$ & 180 & 0.03 (0.55) & 6 \\
AM 1444-302 & 200 & 3.38 (0.97) & 480 \\
AM 1459-722 & 220 & 0.14 (0.80) & 14 \\
NGC 5799 & 240 & 0.41 (0.88) & 33 \\
NGC 5812$^c$ & 180 & 0.10 (0.67) & 21 \\
NGC 5813$^a$ & 180 & 0.09 (0.63) & 18 \\
NGC 5903$^c$ & 180 & 0.09 (0.58) & 18 \\
NGC 6251 & 160 & 5.53 (0.91) & 1810 \\
NGC 6314 & 220 & 4.92 (0.96) & 510 \\
NGC 6482$^a$ & 220 & 0.34 (0.66) & 35 \\
NGC 6483$^c$ & 160 & 0.21 (0.64) & 69 \\
IC 4797$^c$  & 180 & 0.41 (0.72) & 85 \\
NGC 6758$^c$ & 200 & 0.14 (0.62) & 20 \\
IC 4889$^c$  & 200 & 0.18 (0.73) & 25 \\
NGC 6909$^c$ & 180 & 0.08 (0.70) & 16 \\
\hline
\end{tabular}
\begin{minipage}[]{8cm}
\begin{footnotesize}
{\bf Note:} Ionized gas and optical dust mass estimates used for Fig.\ \ref{f:DustWarm} are taken from: a - Goudfrooij et al.\ (1994a, 1994b); b - Dewangan et al.\ (1999); c - Ferrari et al.\ (1999).
\end{footnotesize}
\end{minipage}
\end{table}

\section{Discussion}
\label{S:discuss}
This is our second paper presenting new results from an on-going programme to study the ionized gas and dust in E/S0 galaxies. 
We are motivated by previous ISM surveys revealing ionized gas in nearly every galaxy where dust is found (e.g., Goudfrooij et al.\ 1994a, 1994b; Macchetto et al.\ 1996; Martel et al.\ 2004; Sarzi et al.\ 2010). A general result, which holds also for our sample galaxies, is that the dust and ionized gas almost always strongly correlate while spread over scales from few hundred parsecs to several kpc.
Hence, any theory of the origin and evolution of the ISM needs to account also for the co-existence and co-evolution of these two components with their various phases.

The surprisingly high detection rate of ISM in E/S0 galaxies requires a clarification. 
In late-type galaxies the dust is formed primarily in cold, dusty clouds where it is deposited into the ISM by condensation of thermally-unstable material from the coronal phase or is injected by mass-loss from evolved red giant stars (Dwek \& Scalo 1980).
Although this is also an obvious source of ISM in elliptical galaxies, their overall dust content could be reduced depending on the  grain destruction mechanisms in action.
On the other hand, some material could be accreted onto an elliptical galaxy by tidally capturing a gas-rich spiral or dwarf, or through a collision of two gas-rich galaxies of comparable mass.

To distinguish between internally-produced and externally-acquired ISM considerable efforts were investigated in studying the multiphase gas in E/S0 galaxies.
Here, however, we rely primarily on the properties of the dust to draw conclusions on the origin of the ISM. 
While modern multi-wavelength surveys show that the presence of dust in E/S0 galaxies is the rule rather than the exception, note that the dust distribution is usually concentrated on the nucleus and extends out to radii of a few hundred parsecs, rather than spread throughout the galaxy as in most of our sample galaxies (Sarzi et al.\ 2006; Finkelman et al.\ 2010a) .
Such nuclear dust structures are usually associated with well-defined gaseous rings or discs that are often oriented parallel to the major axis and trace separate stellar components (Krajnovi\'{c} et al.\ 2008; Krajnovi\'{c} et al.\ 2011). The distribution of mean misalignment between the stellar and gaseous angular momenta of these inner structures is inconsistent with a purely external origin, implying a good balance between gas that has been accreted and internally recycled (Sarzi et al.\ 2006). 

Our sample does not pretend to be representative of E/S0 galaxies, but rather of a distinct subclass of E/S0 galaxies that exhibit prominent, large-scale dust features in their optical images. However, since we aim to test whether the tight relationship between gas and dust holds over various spatial scales, we  refer in our study also to E/S0 galaxies with inner, sub-kpc dust components that were previously suggested of being externally accreted (see e.g.\ Martel et al.\ 2004).   

Many of the galaxies in our sample show signs of recent interaction, with NGC 5128 being the most famous example (see Finkelman et al.\ 2010a). 
Searching NED and known catalogues of groups of galaxies (Huchra \& Geller 1982; Geller \& Huchra 1983; Garcia 1993; Tago et al.\ 2010) we find no correlation between the amount of dust and the environment of the galaxies (see also Tran et al.\ 2001; Patil et al.\ 2007). However, Kaviraj et al.\ (2011) did find that the environments of dusty early-type galaxies are typically less dense than those of systems with no sign of optical dust obscuration. Kaviraj et al.\ found also that dusty early-type galaxies are much more likely to host star formation and nuclear activity than early-type galaxies with no prominent dust lanes. 
Furthermore, both the active galactic nucleus (AGN) and starburst ages were found to be younger in dust-lane early-type galaxies than in other early-type galaxies, implying these were triggered by gas-rich galaxy interactions (Shabala et al.\ 2011).
In this view, relaxed gaseous discs and chaotic filamentary structures, and different rates of star formation and black hole accretion rates, could represent different states of the same type of event.
We show below that the grain size distribution, content and spatial distribution of dust in our galaxies support this picture.

\subsection{Grain size distribution}
While the presence of dust in a cold ISM environment of late-type galaxies is natural, hot coronal gas in E/S0 galaxies will erode the dust grains via thermal sputtering, or even destroy them, depending on the grain size.
This will change the dust size distribution and with it the wavelength dependence of the extinction.
Other mechanisms, such as grain-grain collisions, are less efficient in destroying dust in galaxies embedded in hot gas (see Goudfrooij 1999).

\begin{figure*}
\begin{center}
\begin{tabular}{cc}
 \includegraphics[width=8cm]{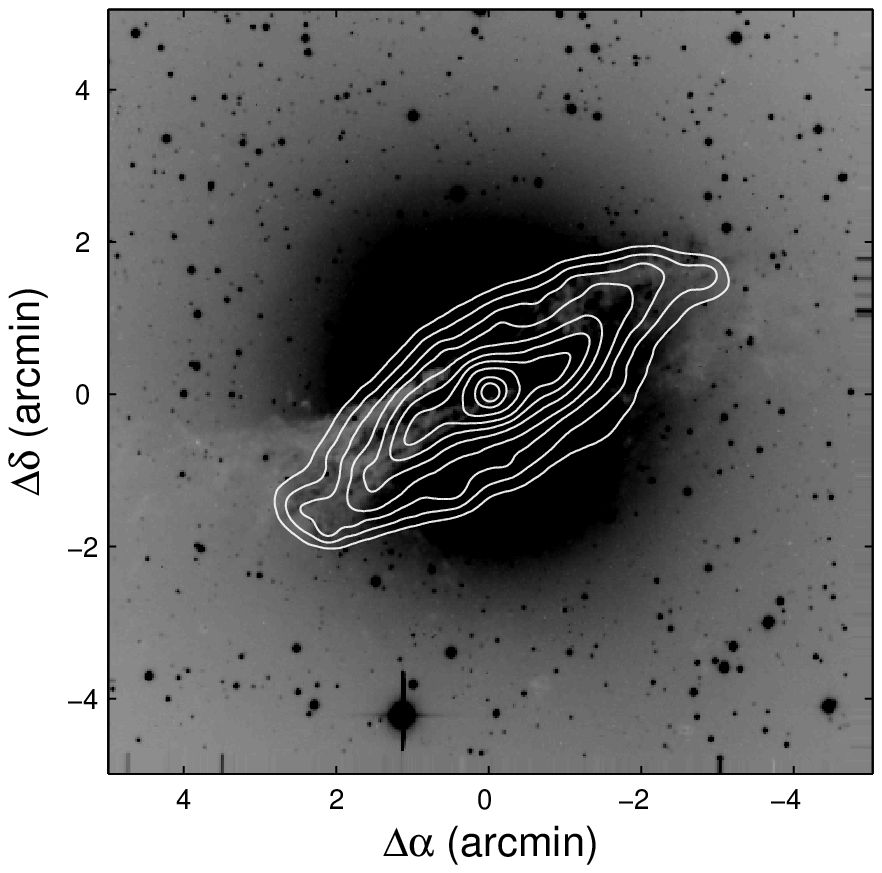} & \includegraphics[width=8cm]{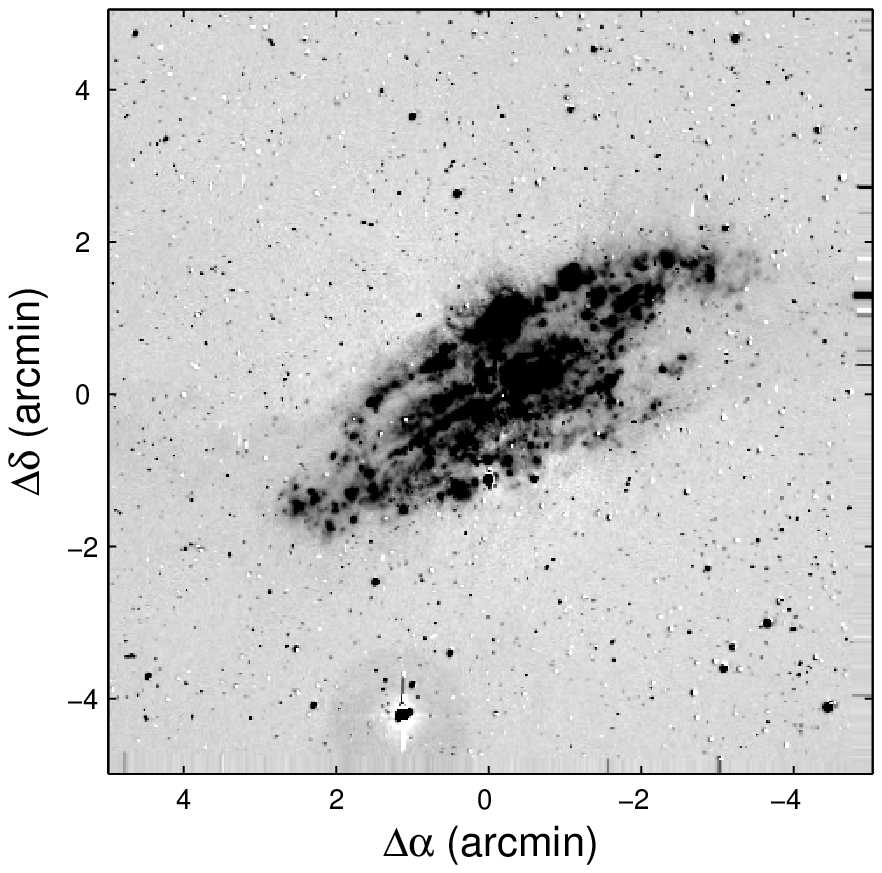} 
\end{tabular}
\end{center}
  \caption{ Left panel: 22$\mu$m map of NGC 5128, shown as contours superposed on our R-band image of the galaxy. 
Right panel: H$\alpha$ continuum-subtracted image of NGC 5128. Both images are displayed as negative. \label{f:NCG5128}}
\end{figure*}

Studying the wavelength dependence of extinction in E/S0 galaxies implies that the dust grain properties are remarkably similar to those of dust grains in the Galaxy with an average grain size of $\sim$0.1 $\mu$m (Finkelman et al.\ 2008; 2010a, 2010b).
If grains condense out of a cooling flow they would be expected to be destroyed within less than $\sim$100 Myr.
However, such a destruction rate cannot account for the grain size or for the total build-up of dust mass observed in this class of galaxies (Patil et al.\ 2007; Casuso \& Beckman 2010; Clemens et al.\ 2010; Kaneda et al.\ 2011). 
While several studies point that the mean grain size is somewhat smaller for galaxies with well-ordered dust lanes (e.g., Goudfrooij et al. 1994; Patil et al 2007), note that sputtering destroys the smallest grains first, hence actually increasing the average grain size.
Furthermore, as demonstrated for NGC 5128 in Figs.\ \ref{f:BBcurves} and \ref{f:NCG5128}, we detected in our sample the presence of very small ($a\lesssim0.01\mu$m) grains by their mid-IR emission and found that their distribution is associated with the optical dust. 
In fact, Fig.\ \ref{f:DustWarm} shows that the masses of these two dust components also correlate; the Spearman rank test shows that the null hypothesis of no correlation has a probability of only about 1\%.
Previous near- and mid-IR studies detected also PAH features in a large fraction of (X-ray emitting) E/S0 galaxies where considerable amounts of dust are clearly present (Xilouris et al.\ 2004; Pahre et al.\ 2004; Kaneda, Onaka \& Sakon 2005; Kaneda et al.\ 2010; Panuzzo et al.\ 2011).
According to these findings, the dust grain distribution seems to be indifferent to the presence of hot gas.

\subsection{Dust content}
A large part of our sample galaxies are not listed in catalogues of X-ray sources, probably due to the relatively low sensitivity of large X-ray surveys of hot gas in elliptical galaxies.
While these shallow surveys paid considerable attention to X-ray bright `giant' elliptical galaxies, deep Chandra and XMM-Newton observations revealed substantial amounts of hot gas in `normal' elliptical galaxies as well (see e.g., Mathews \& Brighenti 2003; Diehl \& Statler 2007; Boroson, Kim \& Fabbiano 2011).
As a general rule, less massive galaxies are expected to be less able to retain the hot gas, which could escape the gravitational potential of the galaxy. However, the hot gas content probably depends to some extent also on the evolutionary history of a galaxy.
For instance, E/S0 galaxies that have undergone a recent merger have low X-ray luminosities 
while relaxed E/S0 galaxies are stronger X-ray emitters (Sansom, Hibbard \& Schweizer 2000). This can be explained if hot gas halos build up through mass loss from stars over a timescale of several Gyr after a merger.\\

Our sample galaxies lie in diverse environments, cover a wide range of size and have probably experienced different merger histories.
It is therefore reasonable to assume that the hot gas content varies significantly between these objects.
On one hand, observations of ionized gas in E/S0 galaxies where x-ray halos were not detected show that at least some of the `warm' ionized gas does stay within the galaxy potential well and is often accompanied by dust (Macchetto et al.\ 1996). 
On the other hand, massive, X-ray bright galaxies are able to retain their warm gas content, while the internally produced dust is subject to rapid erosion by sputtering. Therefore, internally-produced ionized gas is expected to anti-correlate with the dust component.

To test this, we plot in Fig.\ \ref{f:DustHa} the ionized gas mass versus the dust mass, including also data from previous studies (Goudfrooij et al.\ 1994a, 1994b; Macchetto et al.\ 1996; Martel et al.\ 2004). The two components have comparable masses over about 4 orders of magnitude, spreading over spatial scales from the nuclear regions to extended structures several kpc away from the centre. 
The ionized gas spatially correlates not only with the optically obscured regions, but also with the hot dust emission, which suggests that the source of ionization and dust heating is the same (see Figs.\ \ref{f:DustWarm} and \ref{f:NCG5128} here, and figs.\ A1-8 in Finkelman et al.\ 2010a). 
These findings, therefore, provide important evidence against an internal origin of these components.
In fact, since the observed mass relation seems to be independent of the hot gas content of each galaxy, there must be a mechanism to protect dust grains embedded in hot ambient gas from destruction. 

The dust grain lifetime could be much longer for material acquired via an interaction as part of a cooler medium that is (partly) isolated from the hot medium.
Inside dense molecular clouds, dust would be self-shielded against the diffuse galactic UV radiation and the hot thermal electrons in the adjacent hot gas (Temi, Brighenti \& Mathews 2007). 
On the other hand, other ionizing sources, such as AGN or star formation activity, could disperse and heat some of the dusty gas. These conditions could possibly explain why dust is heated but not destroyed, even in the presence of hot plasma.   
Shocks triggered by cloud-cloud interactions during accretion of cold gas could produce also a multiphase dusty gas composed of a dust-free, X-ray emitting plasma and a cold component of mixed dust and molecular gas (Guillard et al.\ 2009). However, whether shock-heating can account for the observed line-emission characteristics is still unclear (Tang et al.\ 2011).
If the gas and dust in the ionized regions are not the product of ISM cooling processes, they could be part of the original warm phase of the accreted ISM. Possible evidence for such a process was given by Kauffmann, Li \& Hickman (2010) who suggested that satellite galaxies are likely to trace an underlying reservoir of ionized gas that can be accreted onto the host galaxy.

Although E/S0 galaxies were once thought to be almost entirely devoid of cold gas, we know by now that the molecular gas in E/S0 galaxies in invariably associated with dust lanes (Wang et al.\ 1992; Sage \& Galletta 1993; Combes et al.\ 2007; Krips et al.\ 2010; Young et al.\ 2011).
A correlation between the cold molecular gas reservoir and the cold dust content was also established for far-IR selected early-type galaxies (see Kaviraj et al.\ 2011 and references therein). However, we caution that a significant part of the far-IR emission originates probably from dust distributed throughout the galaxy and heated by the ambient stellar radiation field (see Section \ref{S:dust_distribution}). This diffuse dust component is associated with the diffuse atomic gas phase rather than with the dense molecular gas.

The molecular gas in E/S0 galaxies does not correlate with the underlying stellar population implying it was most likely externally accreted (Wiklind, Combes \& Henkel 1995). 
Several E/S0 galaxies with dust lanes were also found to contain significant amounts of cold neutral hydrogen, generally distributed in an extended disc-like structure of low surface density (Oosterloo et al.\ 2002; Serra et al.\ 2008). 
While the latter is believed to be a result of an accretion event, the HI-to-dust ratios measured are considerably lower than the gas-to-dust ratio typical of spiral galaxies. It is therefore possible that much of the gas in these galaxies could be in molecular rather than atomic form. 
Since molecular gas normally has a higher surface density than atomic gas, it falls deeper in the galaxy potential well and is more difficult to remove by interactions with neighboring galaxies (Young et al.\ 2011). Part of the atomic gas could also be heated by hot plasma and later cool to the cold neutral medium temperatures, condense and become molecular. Consequently, the cold gas in the central regions could be dominated by molecular gas (Oosterloo et al.\ 2010).

\subsection{Dust distribution}
\label{S:dust_distribution}
Studying NGC 5128 in the far-IR to sub-mm wavelength range Leeuw et al.\ (2002) derived a total dust mass of $2.2\times10^6 \, M_\odot$ within 225 arcsec of the nucleus. This value is consistent with our estimate of the dust mass from optical extinction.
However, far-IR observations of more distant, less-resolved galaxies indicate that the entire dust mass in E/S0 galaxies is generally at least an order of magnitude higher than estimated by optical extinction, as illustrated in Fig.\ \ref{f:ColdDust}.
The discrepancy can be explained by the presence of a diffuse, massive dust component which light attenuation is difficult to detect (Goudfrooij \& de Jong 1995). Significant reservoirs of dust might also be located in the central parts of these galaxies (Temi et al.\ 2007; Leeuw et al.\ 2008).

The distribution of the optical dust is difficult to explain purely by internal processes.
If dust is produced by stellar mass loss, it is expected to follow the stellar light distribution, rather than lie in well-organized structures (see for instance Athey et al.\ 2002; Bregman \& Athey 2004).
AGN outflows disrupting the dense central dusty structure in the galactic core could generate morphologically chaotic or asymmetric obscuring dust features away from the centre, but not well-settled dust discs (Temi et al.\ 2007).
Furthermore, Bregman et al.\ (1998) argued that the observed mass discrepancy between the two dust components cannot be accounted for in hot gas environments if the ISM was internally-generated, unless the grain sizes are typically much higher than for the Galaxy. The authors concluded also that the ratio between the masses derived by IRAS data and optical extinction would appear to grow after a merger.

We note that there seems to be no clear correlation between far-IR emission or dust content and blue luminosity in early-type galaxies (e.g., Forbes 1991; Goudfrooij \& de Jong 1995; Trinchieri \& Goudfrooij 2002; Temi et al.\ 2007; Kaviraj et al.\ 2011; Smith et al.\ 2011).
However, since previous far-IR surveys (e.g., IRAS) were not sensitive to dust colder than $\sim20$ K, testing this correlation and the presence of different dust components requires detecting the thermal emission of the very cold dust phase at sub-mm wavelengths. 
To establish what is the dominant source of dust in E/S0 galaxies requires also mapping its emission for a representative sample of galaxies (Temi et al.\ 2004; Xilouris et al.\ 2004). 
The details of dust grinding processes that could destroy such a correlation should also be accounted for.

Finally, although we showed throughout this paper that the association of gas and dust argues strongly against their internal origin, explaining why externally-acquired dust and ionized gas have comparable masses is far from trivial. 
Considering the various ionization mechanisms in action over different spatial scales makes it difficult to account for such a relation in any scenario (see Sarzi et al.\ 2010). Moreover, if the ISM is acquired from a gas-rich dwarf or through a merger between spiral galaxies, then the overall gas mass is expected to be about 2-3 order of magnitude higher than the overall dust content (e.g., Walter et al.\ 2007; Mu\~{n}oz-Mateos et al.\ 2009). 
Where is the rest of the gas and, assuming that the ionization sources should not be confined to the dust lane, why do we not see significant amounts of dust-free ionized gas?
Solving these issues seems to require making a full inventory of the multiphase ISM ingredients, which is beyond the scope of our study.

\section{conclusions}
\label{S:conclusions}
We present broad-band and narrow-band images to study the content of dust and ionized gas in E/S0 galaxies with dust lanes.
We find that: \\
(1) the grain size distribution typically follows that of the Galaxy;\\
(2) the dust mass, measured from the optical extinction by `large' dust grains, ranges from $10^3$ to $10^7 \, \mbox{M}_\odot$;\\
(3) the hot dust mass, measured from the 22$\mu$m emission from `small' dust grains, ranges from 1 to 3000 $M_\odot$;\\
(4) the cold dust mass, measured from IRAS data, is typically much higher than the optical dust mass.\\
(5) the ionized gas content strongly correlates with that of the optical dust;\\
(6) the ionized gas is morphologically associated with the optical dust structure and the hot dust distribution;

We argue that these observed relations indicate that the ionized gas and the obscuring material have the same origin, are heated by the same sources and are well mixed. 
We conclude that an internal origin of the dust and ionized gas in E/S0 galaxies with dust lanes is highly unlikely;
the hot gas content of E/S0 galaxies is quite heterogeneous and expected to affect differently the grain size distribution, mass content and dust distribution of individual galaxies, whereas our findings are independent of the hot gas content of each galaxy.
We argue also that our results are consistent with the `evaporation flow' hypothesis, albeit with some uncertainty as to the exact details of the process. If the dusty gas that we observe in the optical is part of the dense material arriving from outside during an accretion or merger event, than it could survive destruction even in hot and extreme environments. 
Relaxed gaseous discs and chaotic filamentary structures represent in this picture different states of similar events.
The frequent detection of tidal features, atomic and molecular gas and kinematically decoupled gas components in E/S0 galaxies with dust lanes support this proposed view.

\section{Acknowledgement}
We thank Robert C.\ Kennicutt Jr.\ and Janice C.\ Lee for allowing us to publish their H$\alpha$ image of NGC 5128.
IF wishes to thank Rivay Mor for useful discussions.
SB, AK and PV acknowledge support from the National Research Foundation of South Africa. 

Based on observations with the VATT: the Alice P. Lennon Telescope and the Thomas J. Bannan Astrophysics Facility.

This research has made use of the NASA/IPAC Extragalactic Database
(NED) which is operated by the Jet Propulsion Laboratory,
California Institute of Technology, under contract with the
National Aeronautics and Space Administration. 

This publication makes use of data products from the Wide-field Infrared Survey Explorer, which is a joint project of the University of California, Los Angeles, and the Jet Propulsion Laboratory/California Institute of Technology, funded by the National Aeronautics and Space Administration.

This publication makes use of data products from the Two Micron All Sky Survey, which is a joint project of the University of Massachusetts and the Infrared Processing and Analysis Center/California Institute of Technology, funded by the National Aeronautics and Space Administration and the National Science Foundation.

\end{document}